\documentclass{article} 
\usepackage{iclr2025_conference,times}

\usepackage{amsmath,amsfonts,bm}

\def\eqref#1{equation~\ref{#1}}

\def\1{\bm{1}}

\DeclareMathAlphabet{\mathsfit}{\encodingdefault}{\sfdefault}{m}{sl}
\SetMathAlphabet{\mathsfit}{bold}{\encodingdefault}{\sfdefault}{bx}{n}

\usepackage{graphicx}
\usepackage{hyperref}
\usepackage{url}
\usepackage{algorithm}
\usepackage{algpseudocode}
\usepackage{amsmath}
\usepackage{xcolor}
\usepackage{float}

\title{Accelerated Portfolio Optimization and Option Pricing with Reinforcement Learning}

\author{Hadi Keramati, Samaneh Jazayeri   
\\
Magnative AI\\
\texttt{\{keramati,samanehj\}@magnative.ai} \\
}

\iclrfinalcopy 
\begin{document}

\maketitle

\begin{abstract}
We present a reinforcement learning (RL)-driven framework for optimizing block-preconditioner sizes in iterative solvers used in portfolio optimization and option pricing. The covariance matrix in portfolio optimization or the discretization of differential operators in option pricing models lead to large linear systems of the form $\mathbf{A}\textbf{x}=\textbf{b}$. Direct inversion of high-dimensional portfolio or fine-grid option pricing incurs a significant computational cost. Therefore, iterative methods are usually used for portfolios in real-world situations. Ill-conditioned systems, however, suffer from slow convergence. Traditional preconditioning techniques often require problem-specific parameter tuning. To overcome this limitation, we rely on RL to dynamically adjust the block-preconditioner sizes and accelerate iterative solver convergence. Evaluations on a suite of real-world portfolio optimization matrices demonstrate that our RL framework can be used to adjust preconditioning and significantly accelerate convergence and reduce computational cost. The proposed accelerated solver supports faster decision-making in dynamic portfolio allocation and real-time option pricing.
\end{abstract}

\section{Introduction}
Iterative methods are popular methods for portfolio optimization and option pricing \cite{gulliksson2020iterative}. Option contracts are widely traded and are an inherently asymmetric opportunity that requires price valuation. The first approach for using the discretized Brownian motion for option pricing was suggested by \cite{bachelier1900theorie}. \cite{black1973pricing} suggested a partial differential equation to assess a fair price for European option contracts to consider the time and risk associated with the underlying asset. The Black–Scholes formula and some of the later suggested models have closed form solutions, but numerical methods provide solutions across the discretized time domain and are in most cases faster than a closed form formula in multi-dimension space. Many of these models, such as jump-diffusion models, consider discontinuous moves in the market in addition to the continuous Brownian motion which can be solved by iterative numerical solvers \cite{baldinaiterative,merton1976option}. Large portfolio optimization (e.g. Markowitz mean variance optimization) is also a linear system with a form of $ \mathbf{A} \textbf{x} = \textbf{b}$ where $\mathbf{A}$ is the covariance matrix of asset returns, $\textbf{x}$ are the portfolio weights for each asset, and $\textbf{b}$ is a constraint vector ensuring a fully invested portfolio \cite{bajeux2012krylov,gilli2019numerical}. Directly inverting the matrix $\mathbf{A}$ to solve this portfolio optimization problem is extremely expensive, which is why iterative methods are used to compute the vector $\mathbf{x}$. 

Solving these linear systems of equations that govern portfolio optimization and option pricing with minimal computational overhead is crucial for successful deployment because it involves real-time decision making \cite{gaikwad2009gpu}. For example, portfolio managers must update the weights of each asset or asset class to reduce risk and increase return. However, when the portfolio includes a large number of assets, the covariance matrices become large. Option strategies also require real-time updates in option price valuation, but the corresponding matrices after discretization, mainly using finite difference with fine meshes, become large and time-consuming to solve \cite{ikonen2008efficient,korn2001option}. Solving ill-conditioned problems, particularly in multiasset systems where each asset imposes a specific structure, results in slow convergence. Using preconditioners to accelerate the solvers, particularly for ill-conditioned matrices, is necessary, and block preconditioners take advantage of the system's structure such as asset classes and boundary conditions to accelerate the convergence of the iterative solvers. 

Iterative methods for solving Partial differential equations (PDEs) struggle with nonsymmetric or poorly conditioned problems \cite{greenbaum1996any}. Flexible generalized minimal residual method (FGMRES) is a method that works based on the Krylov subspace and minimizes the residual in this subspace, which is particularly effective for large-scale, sparse problems arising from PDE discretization \cite{saad1986gmres,saad1993flexible}. FGMRES adds flexibility to adapt more effectively to problems where the preconditioner might change or require adjustment during iterations. Preconditioning is a crucial step in iterative methods, as it transforms the original problem into a form that allows for faster convergence. Common preconditioning strategies include techniques like Incomplete LU (ILU) factorization, multigrid methods, and domain decomposition \cite{benzi2002preconditioning}. The block-partitioned preconditioner, which divides the matrix into blocks and applies preconditioning within each block, has shown significant promise in improving the efficiency of both GMRES and FGMRES \cite{saad2003iterative}. In recent years, machine learning approaches have emerged as a means to further optimize iterative solvers. \cite{chen2024graph} used a graph neural network to predict preconditioners for FGMRES. Reinforcement learning (RL), in particular, has shown potential in dynamically tuning solver parameters for improved performance. For example,  \cite{han2018machine} demonstrated the use of RL to optimize parameters in iterative solvers, leading to faster convergence. Proximal Policy Optimization (PPO), a popular RL algorithm, has been successfully applied to adapt preconditioning thresholds in real-time, balancing computational cost with convergence efficiency \cite{schulman2017proximal}.

This paper explores the integration of RL and the block-partitioned preconditioner for portfolio optimization and option pricing problems. Using the dynamic adjustment capabilities of PPO, we train an agent to adjust the size of the preconditioner block during the iterative process to enhance the speed of the solver to achieve an accelerated real-time solver that is effective for portfolio weight adjustment and option pricing over time.  

\section{Preliminaries}

\subsection{Portfolio Optimization}

In mean-variance portfolio optimization, the portfolio weight vector \(\textbf{x} \in \mathbb{R}^n\) minimizes portfolio variance while meeting a specified target return and ensuring full investment. This constrained-based optimization is stated as follows. 

\begin{equation}
\begin{aligned}
\min_\textbf{x}\quad & \frac{1}{2} \textbf{x}^T \Sigma \textbf{x}, \\
\text{s.t.}\quad & \boldsymbol{\mu}^T \textbf{x} = R_{\text{target}}, \\
                & \textbf{e}^T \textbf{x} = 1
\end{aligned}
\end{equation}
where \(\Sigma \in \mathbb{R}^{n \times n}\) is the covariance matrix of asset returns, \(\boldsymbol{\mu} \in \mathbb{R}^n\) is the vector of expected returns, \(\textbf{e} \in \mathbb{R}^n\) is the vector of ones, and \(R_{\text{target}}\) is the desired portfolio return.
To incorporate the constraints into the optimization framework, we form the Lagrangian.
\begin{equation}
\mathcal{L}(\textbf{x}, \lambda_1, \lambda_2) = \frac{1}{2} \textbf{x}^T \Sigma \textbf{x} - \lambda_1 (\textbf{e}^T \textbf{x} - 1) - \lambda_2 (\boldsymbol{\mu}^T \textbf{x} - R_{\text{target}}),
\end{equation}
where \(\lambda_1\) and \(\lambda_2\) are Lagrange multipliers. Taking the derivatives with respect to \(\textbf{x}\), \(\lambda_1\), and \(\lambda_2\) and setting them equal to zero yields the first-order optimality conditions:
\begin{equation}
\begin{aligned}
\nabla_\textbf{x} \mathcal{L}: \quad & \Sigma \textbf{x} - \lambda_1 \textbf{e} - \lambda_2 \boldsymbol{\mu} = 0, \\
\frac{\partial \mathcal{L}}{\partial \lambda_1}: \quad & \textbf{e}^T \textbf{x} - 1 = 0, \\
\frac{\partial \mathcal{L}}{\partial \lambda_2}: \quad & \boldsymbol{\mu}^T \textbf{x} - R_{\text{target}} = 0.
\end{aligned}
\end{equation}
These equations constitute a system of linear equations with unknown vector. 
\[
\textbf{y} = \begin{pmatrix} \textbf{x} \\ \lambda_1 \\ \lambda_2 \end{pmatrix}
\]
Thus, the Karush-Kuhn-Tucker (KKT) conditions can be expressed in the compact form
\begin{equation}
\mathbf{A} \textbf{y} = \textbf{b}
\end{equation}
where the coefficient matrix \(\mathbf{A}\) and the vector on the right side \(\textbf{b}\) are defined by the matrices below. 
\[
\mathbf{A} = \begin{pmatrix}
\Sigma & \textbf{e} & \boldsymbol{\mu} \\
\textbf{\textbf{e}}^T    & 0 & 0 \\
\boldsymbol{\mu}^T  & 0 & 0
\end{pmatrix}, \quad
\textbf{b} = \begin{pmatrix}
0 \\ 1 \\ R_{\text{target}}
\end{pmatrix}
\]
This formulation is a linear system that can be solved using iterative solvers to efficiently find optimal portfolio weights \(\textbf{x}\) in large-scale problems.

\subsection{OPTION PRICING USING FINITE DIFFERENCE METHOD}
In this section, we briefly review the mathematical foundations of option pricing \cite{soleymani2019improved,kovalov2007pricing}. We derive a finite difference discretization that transforms the Black–Scholes PDE into a system of linear equations. We then detail the implicit finite difference scheme that produces a linear system at each time step. For a European option, the value of an option at its expiration is as follows. 
\begin{equation}
V(S, T) = \max(S-K, 0),
\end{equation}
where \(S\) denotes the asset price, \(K\) is the strike price, and \(T\) is the time to expiry. Under the Black--Scholes framework, the option price \(V(S,t)\) satisfies the equation \ref{bs} \cite{chen2017numerical}.
\begin{equation}\label{bs}
\frac{\partial V}{\partial t} + \frac{1}{2}\sigma^2 S^2 \frac{\partial^2 V}{\partial S^2} + rS\frac{\partial V}{\partial S} - rV = 0,
\end{equation}
where \(\sigma\) is the volatility and \(r\) is the risk-free interest rate.

To numerically solve equation \ref{bs}, we discretize both the spatial and temporal domains. We divide the interval \([0, S_{\max}]\) that \(S_{\max}\) is the upper bound of the asset price into \(M\) subintervals of uniform length, $\Delta S = \frac{S_{\max}}{M}$. Therefore, each point in the grid is given by $S_i = i\,\Delta S,\quad i = 0, 1, \ldots, M$. The time interval \([0,T]\) is divided into \(N\) equal steps of size of $\Delta t = \frac{T}{N}$ with intervals $t_n = n\,\Delta t,\quad n = 0, 1, \ldots, N$. For any interior cell we use the following backward approximations \cite{zhao2007compact}.

\begin{equation}
\frac{\partial V}{\partial t} \approx \frac{V_i^n - V_i^{n-1}}{\Delta t},
\end{equation}
where \(V_i^n \approx V(S_i, t_n)\) and \(V_i^{n-1} \approx V(S_i, t_{n-1})\).

The first spatial derivative is approximated by the following difference.
\begin{equation}
\frac{\partial V}{\partial S} \approx \frac{V_{i+1}^{n-1} - V_{i-1}^{n-1}}{2\,\Delta S}.
\end{equation}

Similarly, the second spatial derivative is approximated by
\begin{equation}
\frac{\partial^2 V}{\partial S^2} \approx \frac{V_{i+1}^{n-1} - 2V_i^{n-1} + V_{i-1}^{n-1}}{\Delta S^2}.
\end{equation}

We can substitute the above finite difference approximations into the Black--Scholes PDE evaluated at \(t_{n-1}\).
\begin{equation}
\frac{V_i^n - V_i^{n-1}}{\Delta t} + \frac{1}{2}\sigma^2 S_i^2 \frac{V_{i+1}^{n-1} - 2V_i^{n-1} + V_{i-1}^{n-1}}{\Delta S^2} + rS_i \frac{V_{i+1}^{n-1} - V_{i-1}^{n-1}}{2\,\Delta S} - rV_i^{n-1} = 0
\end{equation}
Rearranging terms for the unknowns results in the following. 
\begin{equation}
-\alpha_i\,V_{i-1}^{n-1} + B_i\,V_i^{n-1} - \gamma_i\,V_{i+1}^{n-1} = V_i^n
\end{equation}
with the coefficients defined by Equations \ref{1}, \ref{2}, and \ref{3}. 
\begin{equation}\label{1}
\alpha_i = \frac{\Delta t}{2}\left(\frac{\sigma^2 S_i^2}{\Delta S^2} - \frac{rS_i}{\Delta S}\right)
\end{equation}
\begin{equation}\label{2}
\gamma_i = \frac{\Delta t}{2}\left(\frac{\sigma^2 S_i^2}{\Delta S^2} + \frac{rS_i}{\Delta S}\right)
\end{equation}
\begin{equation}\label{3}
B_i = 1 + \Delta t\left(\frac{\sigma^2 S_i^2}{\Delta S^2} + r\right)
\end{equation}

Rewriting in matrix terms yields the linear system.
\begin{equation}
\mathbf{A}\,\mathbf{V}^{n-1} = \mathbf{V}^{n}
\end{equation}
where
\[
\mathbf{V}^{n-1} = \begin{pmatrix} V_1^{n-1} \\ V_2^{n-1} \\ \vdots \\ V_{M-1}^{n-1} \end{pmatrix}
\]
and the coefficient matrix \(\mathbf{A}\) is tridiagonal with entries $A_{i,i-1} = -\alpha_i,\quad A_{i,i} = B_i,\quad A_{i,i+1} = -\gamma_i, (i = 1, \ldots, M-1)$.
The boundary conditions at \(S_0\) and \(S_M\), similar to any linear system, are the vector on the right side.

\subsection{Krylov Subspace Solvers}

In this section, we show how linear systems are solved using iterative solvers. We rely on the GMRES algorithm that begins by selecting an initial vector \( v_1 \) and then uses the Arnoldi process to generate an orthonormal basis \(\{\textbf{v}_1, \textbf{v}_2, \ldots, \textbf{v}_k\}\) and an upper Hessenberg matrix \( \mathbf{H}_k \). This process reduces the problem to a least-squares problem that minimizes the residual norm \( \|\mathbf{A}\textbf{x}_k - \textbf{b}\| \). The approximate solution is given by \( \textbf{x}_k = \textbf{x}_0 + \mathbf{V}_k \textbf{y}_k \), where \( \mathbf{V}_k \) is the matrix of basis vectors, and \( \textbf{y}_k \) solves the least squares problem involving \( \mathbf{H}_k \). The Arnoldi process initializes with \( \textbf{v}_1 = \frac{\textbf{r}_0}{\|\textbf{r}_0\|} \), and for each \( j = 1, 2, \ldots, k \), \( \textbf{w} = \mathbf{A} \textbf{v}_j \) is computed. The coefficients \( h_{ij} = \textbf{v}_i^T \textbf{w} \) are calculated for \( i = 1, 2, \ldots, j \), and the vector \( \textbf{w} \) is updated as 
\(
\textbf{w} = \textbf{w} - \sum_{i=1}^{j} h_{ij} \textbf{v}_i.
\)
Finally, the process sets \( h_{j+1,j} = \|\textbf{w}\| \) and normalizes the vector \( \textbf{v}_{j+1} = \frac{\textbf{w}}{h_{j+1,j}} \) to update the solution approximation of $\textbf{x}$ [\cite{Trefethen2022numerical}].

\section{Methodology}

\subsection{Problem Formulation}
The objective of the proposed algorithm is to solve linear systems of equations arising from portfolio optimization or option pricing which are in the form of $\mathbf{A}\textbf{x} = \mathbf{b}$, where $\mathbf{A} \in \mathbb{R}^{n \times n}$ is the known matrix which is usually a sparse matrix and $\textbf{b} \in \mathbb{R}^n$ is a vector. The goal is to find a solution $\textbf{x} \in \mathbb{R}^n$ that minimizes the residual $\mathbf{r} = \mathbf{b} - \mathbf{A}\mathbf{x}$. To improve the convergence rate of the iterative solver, the linear system is preconditioned using the block preconditioner. The matrix $\mathbf{A}$ is split into smaller submatrices of size $k \times k$, resulting in an approximate block diagonal matrix $\mathbf{M}$:

\begin{equation}
    \mathbf{M} = \mathrm{diag}(\mathbf{A}_1, \mathbf{A}_2, \dots, \mathbf{A}_k)
\end{equation}

where $\mathbf{A}_i \in \mathbb{R}^{s \times s}$ are the blocks of $\mathbf{A}$.

\subsection{Block Preconditioner}

Since we are changing the size of blocks of preconditioner in each iteration, we use Flexible GMRES (FGMRES). The Block preconditioner for FGMRES method partitions the  matrix \( \mathbf{A} \) into smaller blocks and applies the preconditioner. This approach is particularly effective for handling large and structured linear systems by improving the condition number of the system matrix, which accelerates the convergence of the iterative solver. In detail, the matrix \( \mathbf{A} \) is divided into \( k \times k \) blocks. For each block, the preconditioning matrix \( \mathbf{M} \) is constructed by performing QR decomposition, which decomposes the block into an orthogonal matrix \( \mathbf{Q} \) and an upper triangular matrix \( \mathbf{R} \). Specifically, for a block \( \mathbf{A}_i \) of \( \mathbf{A} \), the QR decomposition is given by \( \mathbf{A}_i = \mathbf{Q}_i \mathbf{R}_i \). The preconditioner \( \mathbf{M} \) for the entire matrix \( \mathbf{A} \) is then assembled using these decomposed blocks. During the FGMRES iterations, this preconditioner is applied to each block adaptively based on the current residuals. The application of the preconditioner improves the convergence rate by reducing the effective condition number of the system, thus making the iterative process more efficient.

\begin{equation}
    \mathbf{M}_{\mathrm{block}} \approx \mathbf{Q}\mathbf{R}
\end{equation}

where $\mathbf{Q}$ and $\mathbf{R}$ are obtained from the QR decomposition of the corresponding block $\mathbf{A}_{\mathrm{block}}$.

\subsection{Reinforcement Learning for Preconditioning}

A reinforcement learning (RL) agent is used to adjust the preconditioning process. The RL agent is defined by the current residual vector $\textbf{r}$ as the state, the adjustment of the block size for updating the preconditioner as the action, and the negative residual norm as the reward to encourage a reduction in the residual.
The Proximal Policy Optimization (PPO) algorithm is used to train the RL agent to optimize the preconditioning process. As mentioned in the previous section, the GMRES algorithm iteratively refines the solution $\textbf{x}$ as follows:

\begin{equation}
    \textbf{x}_{k+1} = \textbf{x}_k + \textbf{V}_k\textbf{y}_k
\end{equation}

where $\textbf{V}_k$ are the orthonormal basis vectors generated from the Arnoldi process, and $\textbf{y}_k$ is obtained by solving the least-squares problem:

\begin{equation}
    \mathbf{H}_k\textbf{y}_k = \beta\textbf{e}_1
\end{equation}

with $\mathbf{H}_k$ being the upper Hessenberg matrix from the Arnoldi process, $\beta$ the initial residual norm, and $\textbf{e}_1$ the first standard basis vector.

The adaptive preconditioning guided by the RL agent is applied within each GMRES iteration to accelerate convergence. The algorithm terminates when the target residual tolerance is achieved or a maximum number of iterations is reached. The PPO-based block-preconditioner for the FGMRES method improves the traditional approach by integrating a Proximal Policy Optimization (PPO) agent that adjusts the block size \( k \) for the preconditioner and enables it to exploit the underlying structure of the matrix \( \mathbf{A} \) to reduce its condition number. At each iteration, the agent observes the residuals of the linear system and, using its current policy, determines an optimal block size to apply the block-preconditioner. This integer value, which serves as the action taken by the RL agent, dictates how the preconditioner is applied, allowing the method to adapt to the varying structure of the matrix \( \mathbf{A} \) throughout the iterations. Over multiple episodes, the PPO agent learns a policy that effectively configures the preconditioning process to optimize the overall convergence rate. The pseudocode for the framework is presented in Algorithm 1.
\begin{algorithm} 
\caption{Block-partitioned portfolio optimization and option pricing with PPO}
\begin{algorithmic}[1]
\State \textbf{Input:} Matrix \( \mathbf{A} \), vector \( \textbf{b} \), and convergence threshold
\State \textbf{Initialize PPO agent} 
\For{each episode}
    \State Reset environment to $\mathbf{\textbf{x}}_k$
    \While{residual $<$ threshold}
        \State PPO agent choose block size
        \State Initialize preconditioners with new block size and QR decomposition
\State Initialize \( \mathbf{V} \) and \( \mathbf{H} \) for Arnoldi iteration

\For{j in range \( \text{restart} \)}
    \State Apply preconditioner to \( \mathbf{V}[:, j] \) using PPO adjusted block size
    \State Update \( \mathbf{H} \) and \( \mathbf{V} \) via Arnoldi iteration
    \State Solve least squares for \( \textbf{y} \)
    \State Update \( \textbf{x} \) and compute residual
\EndFor
        \State Store transition and update state
    \EndWhile
\EndFor
\end{algorithmic}
\end{algorithm}

During initialization, the sparse matrix $\mathbf{A}$ and the right-hand side vector $\textbf{b}$ are set up, representing the system $\mathbf{A}\textbf{x}=\textbf{b}$. Parameters such as maximum block size, tolerance, iteration limits, and restart frequency are configured. A PPO agent is initialized with the appropriate state and action dimensions, and the number of training episodes is defined. In the training phase, the agent is trained over long episodes to reach the convergence threshold. In each episode, the environment is reset to an initial state and the agent interacts with the environment by selecting actions to adjust the size of the preconditioning block. The PPO agent is trained to use these experiences to refine its value and policy functions. The same process applies to the inference phase, where the preconditioners are initialized by using QR decomposition for each block of the matrix $\mathbf{A}$. The initial guess $\textbf{x}_{0}$ is set to zero, and the initial residual $\textbf{r}_0$ is calculated. During Arnoldi iteration, the preconditioner is applied to the basis vectors $\textbf{V}$. The Arnoldi process updates the matrices $\mathbf{H}$ and $\mathbf{V}$, and a least squares problem is solved to update $\textbf{x}$. Residuals are computed and compared with the convergence criteria, allowing for early termination if the criteria are satisfied. Finally, we return the residuals for each iteration of the PPO-based FGMRES, which can be used to evaluate convergence performance.
\section{Discussion}
\begin{figure}[H]
  \centering
  \includegraphics[width=.7\textwidth]{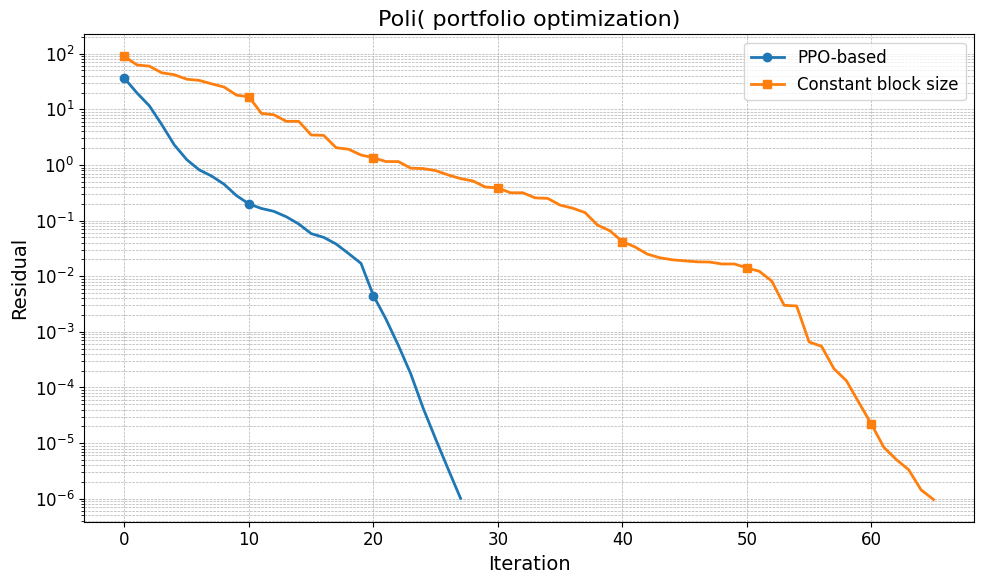}
  \caption{Convergence plots of PPO-based and constant block size for portfolio optimization of the size of 4008}
  \label{fig:poli}
\end{figure}

\begin{figure}[H]
  \centering
  \includegraphics[width=.7\textwidth]{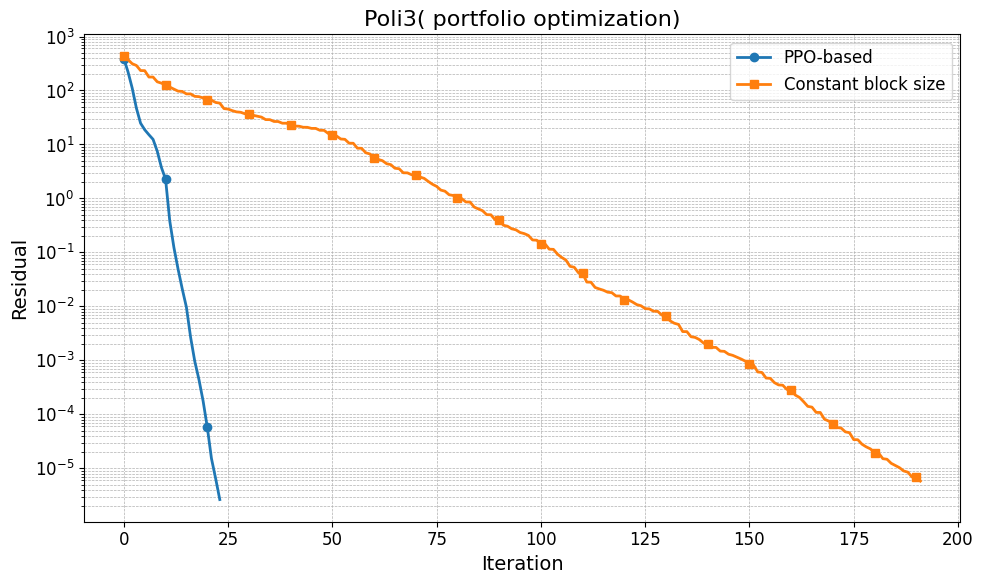}
  \caption{Convergence plots of PPO-based and constant block size for portfolio optimization of the size of 16,955}
  \label{fig:poli3}
\end{figure}

\begin{figure}[H]
  \centering
  \includegraphics[width=.7\textwidth]{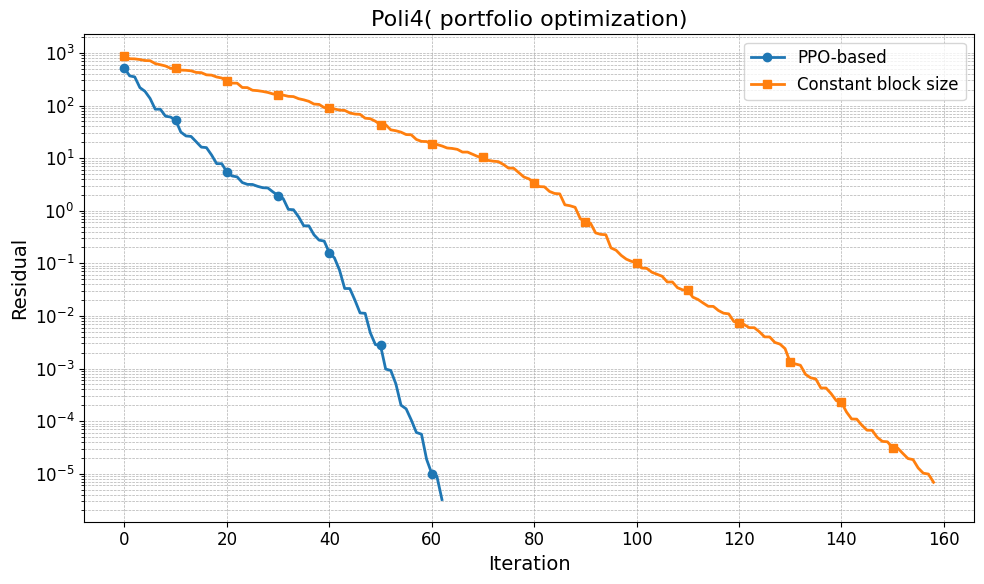}
  \caption{Convergence plots comparing PPO-optimized and constant block-size preconditioning for a portfolio optimization problem with a matrix of size 33,833.
}
  \label{fig:poli4}
\end{figure}

The results of the portfolio optimization matrices \cite{davis2011university} are shown in Figures \ref{fig:poli}, \ref{fig:poli3}, and \ref{fig:poli4}. These real-world portfolio optimization problems are identified by their titles in the matrix collection data, facilitating reproducibility \cite{davis2011university}. Figure \ref{fig:poli} presents the results for a portfolio optimization problem with a matrix of size 4008 that contains 8,188 non-zero elements. In Figure \ref{fig:poli3}, the displayed curve corresponds to a covariance matrix of size 16,955 with 37,849 non-zero elements, which is denser than the matrix shown in Figure \ref{fig:poli}. Figure \ref{fig:poli4} compares the performance of the proposed RL framework with that of a constant block size method for a portfolio whose covariance matrix is of size 33,833 and comprises 73,249 non-zero elements. As mentioned earlier, these matrices are non-symmetric. It can be observed that the pre-trained PPO solver with an adaptive block size converges faster than the solver using a constant block size preconditioner across different matrix sizes. In the PPO-based solver, the block size is limited to the set of integer values considered for the constant block size method, ensuring a fair comparison.

\begin{figure}[H]
  \centering
  \includegraphics[width= \textwidth]{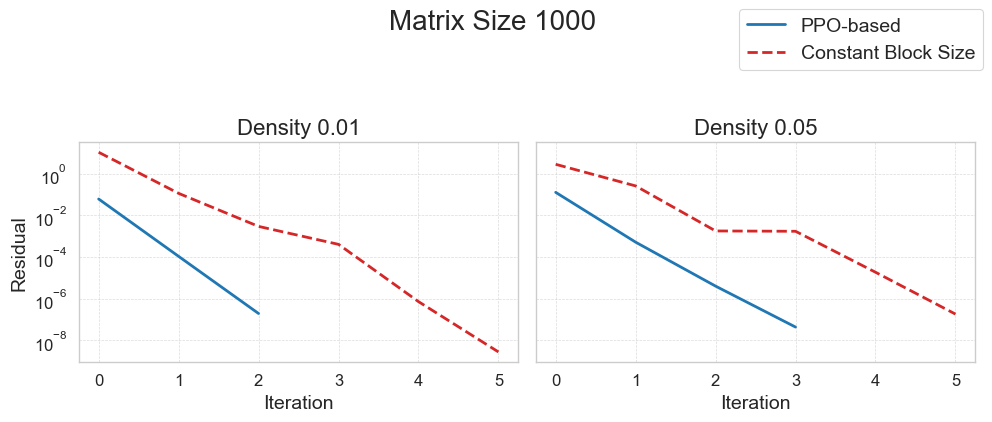}
  \caption{Convergence plots of option pricing for PPO-optimized and constant block preconditioners for a matrix size of 1000}
  \label{fig:option1000}
\end{figure}

\begin{figure}[H]
  \centering
  \includegraphics[width=\textwidth]{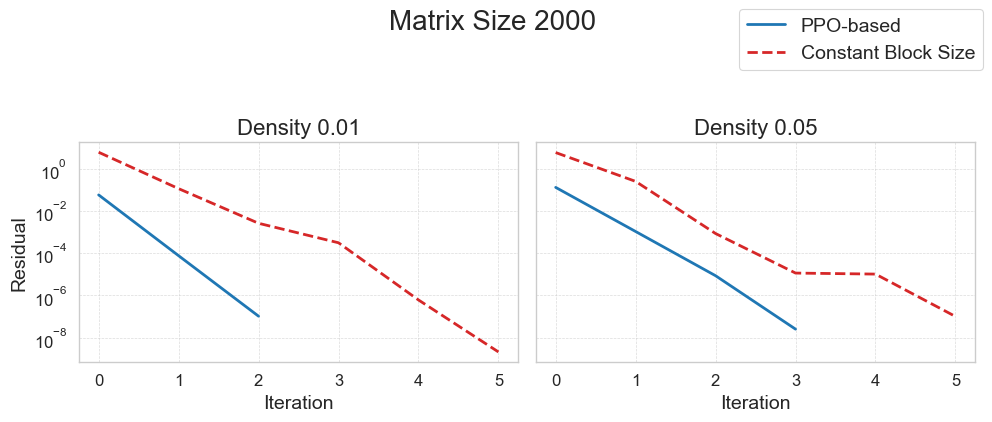}
  \caption{Convergence plots of option pricing for PPO-optimized and constant block preconditioners for a matrix size of 2000}
  \label{fig:option2000}
\end{figure}

Figure~\ref{fig:option1000} presents the convergence behavior of option pricing systems using matrices of size 1000 with varying densities. These matrices are generated from synthetic data with volatilities ranging from 5\% to 25\% and a risk-free rate of 1\%. Each matrix is constructed such that its density, defined as the ratio of nonzero elements to the total number of elements, remains constant. Figure~\ref{fig:option2000} illustrates the convergence of the option pricing system for a coefficient matrix $\mathbf{A}$ of size 2000 at two different densities. In both cases, a clear speedup in obtaining the solution and achieving low residual values is observed across the different matrix sizes and densities. 

Although the computational cost of training the RL agent is nontrivial, the results demonstrate that using an RL agent can significantly reduce the number of iterations required to solve an option pricing problem , in some cases, to as few as two iterations. This reduction in iterations is particularly promising for real-time option pricing applications.

\section{Conclusion}
This paper presents a reinforcement learning framework designed to accelerate the solution of portfolio optimization and option pricing problems formulated as linear systems of equations. The framework constructs a partitioned preconditioner for the FGMRES algorithm and employs an RL agent to learn the optimal block size, thereby enhancing the convergence speed of the solver. Experiments demonstrate accelerated convergence across various matrix sizes and densities for both portfolio optimization and option pricing. These results indicate that the proposed method is especially promising for non-symmetric and ill-conditioned asset models.

\bibliography{iclr2025_conference}
\bibliographystyle{iclr2025_conference}

\end{document}